\documentstyle[aps,prl,epsf,floats]{revtex}

\begin{document}
\draft

\twocolumn[\hsize\textwidth\columnwidth\hsize\csname@twocolumnfalse%
\endcsname

\title  {Roughening transition in a model for dimer adsorption
         and desorption}
\author {Haye Hinrichsen$^1$ and G\'eza \'Odor$^2$\\}
\address{$^1$Max-Planck-Institut f\"ur Physik komplexer Systeme,
         N\"othnitzer Stra{\ss}e 38, 01187 Dresden, Germany}
\address{$^2$ Research Institute for Technical Physics and
         Materials Science,
         P. O. Box 49, H-1525 Budapest, Hungary}
\date    {October 16, 1998}
\maketitle

\begin{abstract}
A solid-on-solid growth model for dimer adsorption and 
desorption is introduced and studied numerically. The special 
property of the model is that dimers can only desorb at the 
edges of terraces. It is shown that the model exhibits a
roughening transition from a smooth to a rough phase. In both phases
the interface remains pinned to the bottom layer
and does not propagate. Close to the transition
certain critical properties are related to those of 
a unidirectionally coupled hierarchy of parity-conserving
branching-annihilating random walks.
\end{abstract}

]


The study of crystal growth and transitions between 
different morphologies of moving interfaces is a field that 
continues to attract great interest~\cite{meakin}.
Various models have been developed in order to describe
the essential features of roughening transitions~\cite{gromod}.
In many cases the critical properties close to the transition 
are universal, i.e., they do not depend on the microscopic 
details of the model. For example, the majority of 
roughening transitions in $d>2$ spatial 
dimensions belongs to the Kardar-Parisi-Zhang 
universality class~\cite{kapezt}. Therefore it is an important 
theoretical task to categorize the possible universality 
classes for roughening transitions in a given geometry. 

It is well known that one-dimensional interface models 
with short-range interactions at thermal equilibrium do not 
exhibit roughening transitions. However, under {\it nonequilibrium} 
conditions such a transition may occur even in one spatial dimension. 
Known examples include polynuclear growth (PNG) models~\cite{pngmod},
solid-on-solid models with evaporation at the edges of
terraces~\cite{monomr}, and certain models for
fungal growth~\cite{fungus}. The key feature of all these 
growth processes is a close relationship to 
directed percolation (DP)~\cite{dirper,regfth}. 
More precisely, the DP process emerges at a particular 
reference height of the interface.  
In the models of Ref.~\cite{monomr} this
reference height is the spontaneously selected bottom layer 
of the interface, whereas in polynuclear and fungal models, 
which evolve by parallel dynamics, the reference level 
propagates at maximal velocity. The sites where 
the interface touches the reference height 
correspond to the active sites of DP. Therefore, in the 
active phase of DP the interface fluctuates close 
to the reference level so that the interface
is smooth. On the other hand, in the inactive 
phase of DP, the interface detaches from
the reference level and evolves into a rough state.
More recently it has been shown that the critical behavior 
at the first few layers next to the reference height
is equivalent to that of unidirectionally coupled 
DP processes~\cite{couple}.

DP itself is the generic universality class for phase transitions 
into  absorbing states and covers a wide range of models. Only a 
few exceptions from DP are known. One of them is the so-called 
{\em parity-conserving} (PC) class~\cite{abmod,barwe,nekim,imdmod,bassl,haye}
which is represented most prominently by branching-annihilating random walks 
with even number of offspring (BAWE)~\cite{barwe}. 
The PC class also includes models with two symmetric absorbing 
states~\cite{nekim,bassl,haye}, where the kinks between different absorbing 
domains may be interpreted as walkers whose number is preserved modulo~2.
In other models~\cite{abmod,imdmod} the two symmetric
absorbing states emerge as checkerboard-like configurations
of particles at even or odd sites, respectively. 

Recently Park and Kahng~\cite{park} posed the question 
whether it is possible to replace the underlying DP mechanism 
in the interfacial growth models of Refs.~\cite{pngmod,monomr,fungus}
by a PC mechanism. To this end they introduced a model which involves 
two symmetric particle species and  exhibits a roughening 
transition in 1+1 dimensions. Since the dynamical rules
at the bottom layer mimic a contact process with two symmetric
absorbing states~\cite{haye}, the authors expected 
the roughening transition to be related to 
the PC class, in the same way as the transition 
in the monomer models of Ref.~\cite{monomr} is
related to DP. But surprisingly the numerical estimates for the 
critical exponents deviated significantly from the 
PC values. Park and Kahng argued that the 
unexpected behavior may be related to the fact 
that dynamical processes at lower levels are strongly 
suppressed by particles at higher levels. In particular,
kinks between different particle species may become frozen 
when they are covered by another layer of atoms.

In this Letter we introduce a model for dimer 
adsorption and desorption where this
suppression effect is much less important. 
As dimers consist of two atoms, the number of particles at each 
height level is preserved modulo $2$. The dynamical rules 
are defined in a way that they mimic a BAWE
at the bottom layer of the interface. It turns out that 
some of the critical properties of the roughening transition 
in this model are indeed characterized by PC exponents. 

\paragraph{Definition of the model.}

%
%
\begin{figure}
\epsfxsize=85mm
\centerline{\epsffile{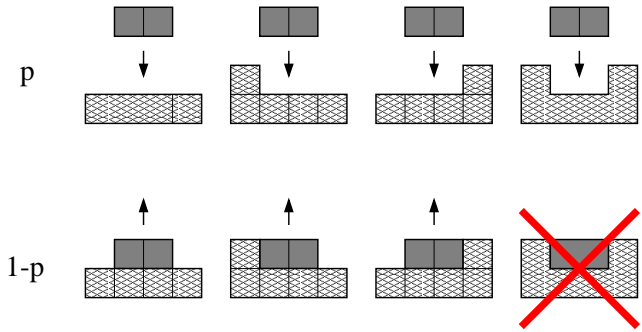}}
\caption{
\label{FigRules}
Dynamical rules: Dimers are adsorbed with probability $p$
and desorbed at the edges of terraces 
with probability $1-p$.
}
\end{figure}

The model is defined on a $d$-dimensional square
lattice with $L^d$ sites and periodic boundary conditions.
The interface height at site $i$ is represented
by an integer height variable $h_i$ which obeys the 
restricted solid-on-solid (RSOS) condition
\begin{equation}
\label{restriction}
|h_i-h_j| \leq 1\,,
\qquad |i-j|=1\,.
\end{equation}
The model evolves by {\it random sequential} updates
according to the following dynamical rules (see Fig.~\ref{FigRules}).
In each attempted update two adjacent sites
$i_1$ and $i_2$ are selected at random. If the heights
$h_{i_1}$ and $h_{i_2}$ are equal, one of the following
moves is carried out. Either dimers are adsorbed
with probability $p$
\begin{equation}
\label{Adsorption}
h_{i_1} \rightarrow h_{i_1}+1
\,, \qquad
h_{i_2} \rightarrow h_{i_2}+1
\ ,
\end{equation}
or dimers desorb with probability $1-p$
\begin{equation}
\label{Desorption}
h_{i_1},h_{i_2} \rightarrow \min_{k \in \langle {i_1},{i_2} \rangle} h_k \,,
\end{equation}
where $\min_{k \in \langle i_1,i_2 \rangle} h_k$ denotes the
minimum height of sites $i_1$ and $i_2$ and their nearest neighbors.
An update will be rejected if it leads to a violation of
the restriction~(\ref{restriction}). 

The above model is defined in arbitrary spatial dimensions $d$
and is translationally invariant in space, time as well as in
height direction. Moreover, it may be easily generalized to 
$n$-particle objects (monomers, dimers, trimers, etc.). 
In particular, for $n=1$ it reduces to the previously studied
monomer growth model of Ref.~\cite{monomr} which is related to DP. 
We use random sequential updates in order to demonstrate that
in contrast to PNG models~\cite{pngmod,fungus},
the phase transition in our model does not require synchronous updates.
We have also studied another variant of the dimer model with 
parallel updates, as will be described elsewhere~\cite{future}.
This cellular automaton model resembles a layer-by-layer 
crystal growth process and exhibits the same universal 
properties as the random-sequential variant.

\paragraph{Phenomenology of the roughening transition.}
%
If the adsorption rate $p$ is very small,
only a few dimers are adsorbed, 
staying for a short time, before they quickly
evaporate again. Therefore, the interface is 
smooth and anchored to the bottom
layer (the spontaneously selected line 
of minimal height).
As $p$ increases, more and more dimers cover 
the surface and large islands with several layers stacked 
on top of each other are formed. When $p$ exceeds
a certain critical value~$p_c$, the mean size of the islands
diverges and the interface heights become uncorrelated 
over long distances, i.e., the interface evolves into a rough state.

In analogy to the monomer models of Ref.~\cite{monomr}
we may expect the interface to detach from the bottom layer
in the rough phase, resulting in a finite propagation velocity.
However, it turns out that in the present case the interface remains
pinned to the initial height and does {\em not} propagate
at constant velocity. This is due to the fact that a
stochastic deposition process cannot create a
dense packing of dimers. The emerging configurations are 
rather characterized by a certain density of
defects (solitary sites at the bottom layer) where dimers cannot be
adsorbed. Because of the RSOS condition~(\ref{restriction}) 
these defects act as `pinning centers'
which prevent the interface from growing. 
The pinning centers cannot disappear spontaneously,
rather they can only diffuse by interface
fluctuations and recombine in pairs so that their
number is expected to decrease extremely slowly.
Therefore, the interface of an infinite system in the rough phase
does not propagate at constant velocity. Instead the average 
height increases only logarithmically with time,
as will be shown below.

In order to illustrate the PC mechanism in this model, 
let us interpret the sites at the bottom layer
$h_i=0$ as $A$-particles. Adsorption and desorption processes
correspond to certain effective reactions of the $A$-particles. 
For example, the adsorption of dimers at the bottom layer 
corresponds to a  pair-annihilation 
process $2A\rightarrow \O$ at rate~$p$. 
On the other hand, when dimers evaporate, two $A$-particles
are created. However, since dimers can only evaporate at the
edges of terraces, the presence of another neighboring 
$A$-particle is required, giving rise to an effective reaction
$A\rightarrow 3A$ at rate $1-p$. These two processes compete 
one another and realize a BAWE. This mapping is, of course, not
exact since the $A$-particles are also coupled to the dynamical 
processes at higher levels of the interface. 
Nevertheless it seems that this feedback does not affect the 
critical behavior of the bottom layer.

\paragraph{Critical properties of the interface width.}

%
%
\begin{figure}
\epsfxsize=85mm
\centerline{\epsffile{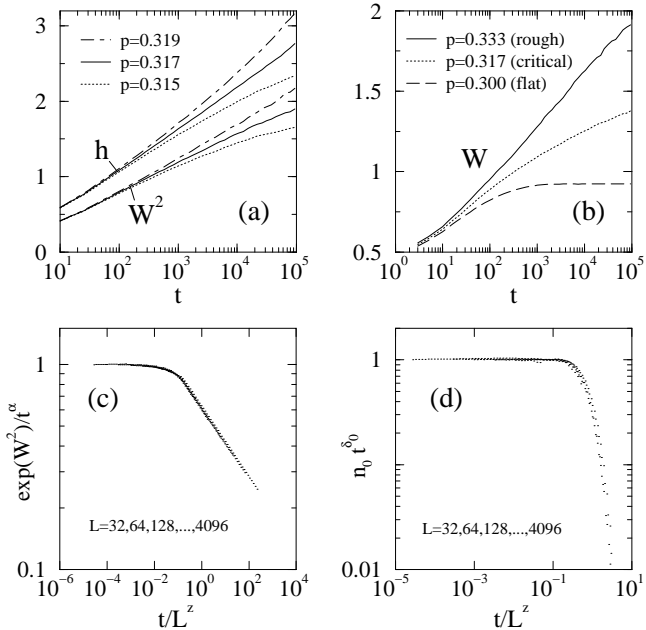}}
\caption{
\label{FigWidth}
Critical properties of the interface. 
In parts (a) and (b) 
the (squared) width and the average
height are shown as functions of time
near and far away from criticality, respectively. 
Parts (c) and (d) show data collapses for
finite-size simulations (see text).
}
\end{figure}
The order parameter that describes a roughening transition 
is the interface width
\begin{equation}
W(t) =
\Bigl[
\frac{1}{L^d} \, \sum_i \,h^2_i(t)  -
\Bigl(\frac{1}{L^d} \, \sum_i \,h_i(t) \Bigr)^2 
\Bigr]^{1/2}
\,.
\end{equation}
In order to investigate its critical properties,
we perform Monte-Carlo simulations in $d=1$ dimensions 
on a lattice with $L=4\,096$ sites up to almost $10^5$ time steps. 
Starting from an initially flat interface $h_i=0$ we find the following
results. Below the critical threshold $p<p_c$ the width saturates at some
finite value which means that the heights $h_i$ are correlated
over long distances. At the roughening transition $p=p_c \simeq 0.317(1)$ 
we observe that the squared width and the average 
height grow {\it logarithmically} 
with time as (see Fig.~\ref{FigWidth}a)
\begin{equation}
W^2(t) \sim \bar{h}(t) \sim \log t \ .
\end{equation}
In the rough phase $p>p_c$ this temporal behavior 
crosses over to $W(t) \sim \log t$
(see Fig.~\ref{FigWidth}b). 

Assuming that the roughening transition is governed by an underlying
PC transition, it is near at hand to conjecture 
the finite-size scaling form
\begin{equation}
\label{WidthScaling}
W^2(L,t) \simeq \log \Bigl[ t^{-\alpha}
\, F(t/L^z) \Bigr]\, ,
\end{equation}
where $F$ is an universal scaling function and 
$z=\nu_\parallel/\nu_\perp= 1.76(5)$ is the dynamical 
scaling exponent of the PC class. The exponent
$\alpha$ is unknown and plays the role
of a roughening exponent.
In order to verify this scaling form, 
we perform finite-size simulations 
at criticality and plot $\exp(W^2)/t^\alpha$ 
for $10^2 \leq t \leq 10^5$ against  $t/L^z$. 
Using the estimate $\alpha=0.172(10)$ 
we obtain a fairly accurate data collapse
(see Fig.~\ref{FigWidth}c), supporting that the 
dynamical exponent in this model is the same
as in PC transitions.

\paragraph{Critical properties of the first few layers.}

%
%
\begin{figure}
\epsfxsize=85mm
\centerline{\epsffile{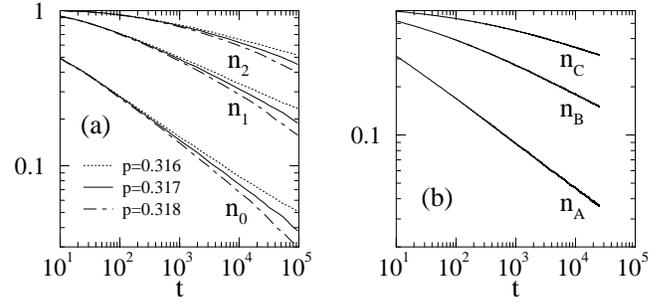}}
\caption{
\label{FigFirstFew}
(a) Densities $n_k$ at the first three levels
as a function of time at and close to criticality. 
(b) Particle densities in unidirectionally coupled
BAWE's at criticality.
}
\end{figure}
We now turn to the critical behavior at the first few layers
of the interface. Let us denote by
\begin{equation}
n_k = \frac{1}{L^d} \sum_{j=0}^{k} \sum_i \, \delta_{h_i,j}
\ , \qquad
k=0,1,2,\ldots
\end{equation}
the density of sites $i$ with $h_i \leq k$.
In analogy to Ref.~\cite{monomr} we expect $n_k$ to
obey the finite-size scaling form
\begin{equation}
n_k(L,t) \sim t^{-\delta_k} \, G(t/L^z) 
\ ,
\end{equation}
where $G$ is a universal scaling function.
$\delta_0 = \delta = \beta / \nu_\parallel$ denotes the usual 
cluster survival exponent of the PC class while the other exponents
$\delta_1,\delta_2,\ldots$ should be independent and smaller than $\delta_0$. 
Starting from a flat interface at $h=0$ in a large system 
with $4\,096$ sites, we measure $n_0$, $n_1$ and $n_2$ at criticality 
(see Fig.~\ref{FigFirstFew}a) which should decay as $n_k(t)\sim t^{-\delta_k}$.
Averaging the slopes over the last decade in time we obtain the
estimates
\begin{equation}
\label{exponents}
\delta_0=0.280(10)
\, , \
\delta_1=0.200(15)
\, , \
\delta_2=0.120(15)
\, .
\end{equation}
The estimates for $\delta_0$ and $z$ can also be verified by a 
finite-size data collapse (see Fig.~\ref{FigWidth}d). 
The numerical value of $\delta_0$ is in fair agreement with
the PC exponent $\delta=0.285(5)$, confirming that the dynamical
processes at the bottom layer belong to the PC universality class. 

\paragraph{Unidirectionally coupled BAWE's.}

\begin{figure}
\epsfxsize=85mm
\centerline{\epsffile{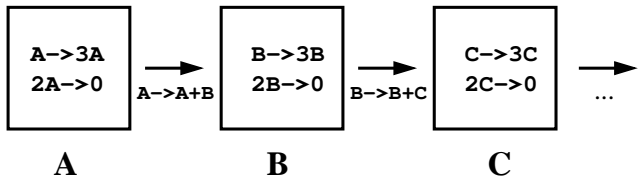}}
\vspace{2mm}
\caption{
\label{FigCoupledScheme}
Unidirectionally coupled BAWE's.
}
\end{figure}

In order to explain the exponents $\delta_1$ and $\delta_2$ 
we propose that the critical behavior at the first few layers 
in the dimer model should be equivalent to that
of a hierarchy of unidirectionally coupled PC processes, 
just in the same way as the monomer model of Ref.~\cite{monomr} 
is related to unidirectionally coupled DP processes~\cite{couple}.
To this end we extend the particle interpretation:  
assuming that $h=0$ is the 
bottom layer of the interface, let us interpret 
sites $i$ with $h_i\leq 0,1,2,\ldots$ 
as particles $A,B,C,\ldots$, respectively. By definition,
particles of different species are allowed to occupy the same 
site simultaneously.
As shown before, the temporal evolution of the $A$-particles resembles a BAWE.
Similarly, the $B$-particles perform an effective BAWE on top of
inactive islands of the $A$-system. Clearly, the 
temporal evolution of the $B$-particles depends strongly on the dynamics of
the $A$-particles. For example, $A$-particles instantaneously create
$B$-particles at the same site, which may be interpreted as an effective
reaction $A \rightarrow A+B$ at infinite rate.
As this reaction does not modify the configuration of the $A$-particles, 
it couples the two subsystems only in one direction without
feedback. On the other hand, the RSOS condition~(\ref{restriction}) 
introduces an effective feedback so that the $A$-particles are 
not completely decoupled from the $B$-particles. However, 
we will assume that in contrast to Ref.~\cite{park} this feedback 
is irrelevant for the critical behavior and can be neglected. 
Similarly, the $C$-particles are coupled 
to the $B$-particles by the effective reaction $B \rightarrow B+C$.
Therefore the dimer model resembles a linear sequence 
of BAWE's which are effectively coupled without 
feedback in one direction, as sketched in Fig.~\ref{FigCoupledScheme}. 

To verify this conjecture, we study three unidirectionally 
coupled BAWE's by Monte Carlo simulations 
(details will be published in~\cite{future}).
We implement the coupling between the systems
in a way that $A$-particles instantaneously 
create $B$-particles at the same site in the $B$-system,
provided that this site is empty. 
Similarly, $B$-particles create
$C$-particles at the same site.
We simulate a hierarchy of three coupled
BAWE's on a lattice of $2\,500$ sites up to
$25\,000$ time steps and measure the densities
of $A$, $B$, and $C$-particles as functions of
time averaged over $1\,000$ independent realizations
(see Fig.~\ref{FigFirstFew}b). 
It turns out that the particle densities 
$n_A,n_B,n_C$ decay in the same way as $n_0,n_1,n_2$
in the dimer model. Averaging the slopes over the
last decade in time we obtain the exponents $\delta_A = 0.280(5)$, 
$\delta_B = 0.190(10)$, and $\delta_C = 0.120(15)$,
which are in fair agreement with the corresponding
exponents of the dimer model in Eq.~(\ref{exponents}).
This supports the hypothesis that the dimer model and 
coupled BAWE's exhibit the same type of critical behavior.

It should be noted that the densities at higher levels do not
scale perfectly. Rather the curves for $k \geq 2$ in Fig.~\ref{FigFirstFew}
appear to be slightly bent, indicating a deviation from perfect scaling.
This curvature is neither related to errors in the estimation of $p_c$
nor to finite-size effects. Similar deviations were also observed in the 
case of DP-related growth models~\cite{couple} and are presumably caused 
by certain infrared divergent diagrams of the underlying field theory.

To summarize, we have introduced an interface model for dimer
adsorption and desorption and have presented numerical evidence 
that the roughening transition in this model may be associated with the
universality class of unidirectionally coupled PC processes.

\vspace{2mm}
Acknowledgements:
The simulations were performed partially on the FUJITSU AP-3000
and System-V parallel supercomputers. G. \'Odor gratefully acknowledges
support from the Hungarian research fund OTKA (Nos. T025286
and T023552) .


\end{document}